\documentclass[11pt,english,times]{article}
\usepackage{amsmath}
\usepackage{amssymb}
\usepackage[english]{babel}
\usepackage{graphicx} % Allows including images
\usepackage{booktabs} % Allows the use of \toprule, \midrule and \bottomrule in tables
\usepackage{color}
\usepackage{graphicx}
\graphicspath{ {images/} }

\author{J\"urg Fr\"ohlich\footnote{Theoretical Physics, ETH Zurich, Wolfgang Pauli Strasse, 8093 Zurich, Switzerland; Email: juerg@phys.ethz.ch}}

\title{Chiral Anomaly, Topological Field Theory, and Novel States of Matter\footnote{Dedicated to the memory of Ludvig Dmitrievich Faddeev -- a great scientist who will be remembered. To appear in: ``Ludwig Faddeev Memorial Volume: A Life in Mathematical Physics'', edited by Molin Ge, Antti Niemi, Kok Khoo Phua, and Leon A Takhtajan (World Scientific, 2018); http://www.worldscientific.com/worldscibooks/10.1142/10811} }

\begin{document}
\maketitle
\begin{abstract}
{\it Starting with a description of the motivation underlying the analysis presented in this paper and a brief survey 
of the chiral anomaly, I proceed to review some basic elements of the theory of the 
quantum Hall effect in 2D incompressible electron gases in an external magnetic field, (``Hall insulators''). 
I discuss the origin and role of anomalous chiral edge currents and of anomaly inflow in 2D insulators with explicitly or spontaneously broken time reversal, i.e., in Hall insulators and ``Chern insulators''. The topological Chern-Simons action yielding the large-scale response equations for the 2D bulk of such states of matter is displayed. A classification of Hall insulators featuring quasi-particles with abelian braid statistics is sketched.\\
Subsequently, the chiral edge spin currents encountered in some time-reversal invariant 2D topological 
insulators with spin-orbit interactions and the bulk response equations of such materials are described.\\
A short digression into the theory of 3D topological insulators, including 
``axionic insulators'', follows next.\\
To conclude, some open problems are described and a problem in cosmology related to axionic insulators is mentioned.\\
As far as the quantum Hall effect and the spin currents in time-reversal invariant 2D topological insulators 
are concerned this review is based on extensive work my collaborators and I carried out in the early 1990's.}

\end{abstract} 
 
\section{General goals of analysis}\label{Intro}

The purpose of this paper is to review some applications of the chiral anomaly and of topological field theory to condensed matter physics, more specifically to the theoretical analysis of novel, topologically protected states of matter. This is an area that has been of much interest to me, ever since the late 1980's, and to which I have made contributions that I feel are interesting and important. For this reason, it makes sense to highlight them, one more time, trying to put the emphasis on some of the most important concepts and most elegant ideas. The results sketched in the following are based on the papers \cite{FM, FK, FZ, FS, FT, FST} (and references given therein) all published in the early nineties. \\
I am confident that the material exposed here would have interested Ludvig Faddeev. Among many other important contributions he greatly enhanced our mathematical understanding of quantum gauge theories and of the origin and implications of the chiral anomaly. Some of his results on gauge theories and anomalies provide the theoretical underpinning for much of what I will discuss in the following.\\

I start by sketching the general motivation underlying the efforts of my collaborators and myself (see \cite{FS, FT, FST, ACF, FP}) that have led to the results discussed in the following.\\ 

\textit{Key purposes} of our work are:
\begin{itemize}
\item {To classify bulk- and surface states of systems of condensed matter, using 
concepts and results from gauge theory, current algebra and General Relativity. In particular, \textit{effective 
actions} ($=$ generating functionals of connected current Green functions yielding expressions for transport 
coefficients, such as conductivities, via Kubo formulae) are used extensively in our analysis; their form being constrained by \textit{gauge-invariance}, by the cancellation of \textit{gauge anomalies} between bulk- and surface terms (``holography''), and by \textit{locality, unitarity} and \textit{``power counting''}. It turns out that these general principles \textit{completely determine} the most relevant terms in the effective actions of, for example, insulators. Using effective actions, one discovers new ways of characterizing novel states of matter, as indicated next.}

\item{To complement the \textit{Landau Theory} of Phases and Phase Transitions
by a \textit{``Gauge Theory of Phases/States of Matter''}, with the purpose of describing novel ``topologically protected'' states of matter that cannot be characterized by a local order parameter.
}
\item{To apply our ``Gauge Theory of Phases of Matter'' to extend or develop:

\begin{itemize}
\item{The theory of the Fractional Quantum Hall Effect; (our work extended over the period from 1989 -- 2012)}

\item{The theory of time-reversal invariant Topological Insulators and of Topological Superconductors; (our first paper appeared in 1993)}

\item{A description of higher-dimensional cousins of the Quantum Hall Effect, with applications in cosmology, e.g., towards
understanding the origin of intergalactic primordial magnetic fields in the Universe, etc.; (1999 -- the present)}
\end{itemize}
}
\end{itemize}

\section*{The chiral anomaly}
While vector currents, $J^{\mu}$, are usually conserved, i.e., $\partial_{\mu}J^{\mu}=0$, \textit{axial} currents, $J^{\mu}_{5}$, defined on space-times of even dimension fail to be conserved, i.e., tend to be \textit{anomalous}, when coupled to gauge fields -- even if they are associated with {\em massless} matter fields. This phenomenon is called \textit{``chiral anomaly''}; (see \cite{GTW} and references given there). \\[3pt]
Let $\vec{E}$ denote the electric field and $\vec{B}$ the magnetic field (or -induction).
In two space-time dimensions, the chiral anomaly is expressed by
\begin{equation*}
\partial_\mu J^\mu_5=\frac{\alpha}{2\pi} E,\quad \alpha:=\frac{q^2}{\hbar},
\end{equation*}
where $q$ is the electric charge of a massless field that gives rise to the current $J^{\mu}_5$, and $\hbar$ is Planck's constant, and by the (non-vanishing) anomalous equal-time commutator
\begin{equation*} 
[J^0_5(\vec{y},t), J^0(\vec{x},t)]\mathop{=} i\alpha\delta'(\vec{x}-\vec{y}).
\end{equation*}
In four space-time dimensions, the anomaly is expressed by
\[
\partial_\mu J^\mu_5=\frac{\alpha}{\pi}\vec{E}\cdot\vec{B},
\]
and the anomalous commutator
\[
[J^0_5(\vec{y},t),J^0(\vec{x},t)]\mathop{=} i \frac{\alpha}{\pi} 
\big(\vec{B}(\vec{y},t)\cdot\nabla_{\vec y}\,\delta\big)(\vec{x}-\vec{y}).
\]
For a massive matter field, there are additional terms contributing to the divergence, $\partial_{\mu}J^{\mu}_{5}$, of the axial  current that are proportional to the mass, $m$, of that field.\\
It is then obvious that the \textit{chiral} currents
$$J_{\ell}^{\mu}:= J^{\mu} - {J_{5}}^{\mu}, \quad \text{and  }\quad  J_{r}^{\mu}:= J^{\mu} + {J_{5}}^{\mu}$$
are anomalous, too, (the signs of their divergence being opposite to one another). \\
In this paper, we will not make use of anomalous commutators. But they play an important role in the study of transport in dissipationless one-dimensional and three-dimensional conductors; see \cite{ACF, FP}.\\
Faddeev and collaborators \cite{FaJa} contributed important insights to a clear understanding of the mathematics underlying the chiral anomaly and anomalous commutators.

\section*{Contents, Credits and Acknowledgements}
This paper is devoted to reviewing efforts to understand novel states of matter that I have personally been involved in and that have spanned more than two decades. I think they have yielded a number of important and elegant results. Of course, numerous other people have contriubuted important insights towards clarifying issues discussed in this paper, and it is completely impossible to do justice to all their contributions in a short review like this one. Some of the results on the fractional quantum Hall effect described in Section 2 overlap with independent ones due to Xiao-Gang Wen \cite{Wen}.\footnote{I will not review work on the integer quantum Hall effect in 2D non-interacting electron gases, and I refrain from quoting papers on this effect, because there are too many of them, and they are well known.} Ideas and results concerning 2D time-reversal invariant topological insulators related to some of the ones described in Section 3 of this paper have also been brought forward by Shoucheng Zhang and collaborators and will be referred to later.\footnote{However, their work only appeared several years after the publication of the one of my collaborators and myself.} In Section 4, some results on 3D time-reversal invariant topological insulators, including axionic insulators, are reviewed. These results, too, overlap with ones described by Shoucheng Zhang and collaborators. 

I am indebted to the following colleagues and friends for collaboration and/or many illuminating discussions on the subject matter covered in this review and for encouragement:\\ 
A. Alekseev, S. Bieri, A. Boyarsky, V. Cheianov, G. M. Graf, B. I. Halperin, T. Kerler, I. Levkivskyi, L. Molenkamp, G. Moore, B. Pedrini, O. Ruchayskiy, C. Schweigert, E. Sukhorukov, J. Walcher, Ph. Werner, P. Wiegmann, and A. Zee.\\
 Very special thanks are due to R. Morf, who introduced me to the quantum Hall effect and joined me for countless discussions, and to U. M. Studer and E. Thiran, without whose devoted help and collaboration my efforts to understand aspects of the fractional quantum Hall effect would have failed.

\section{\mbox{Theory of the Fractional Quantum Hall Effect:}\\ Anomalous Chiral Edge Currrents, Effective\\
Actions, Classification of Hall Fluids}\label{Sec1}

\section*{Setup, basic quantities}

We consider a two-dimensional electron gas (2D EG) forming at the interface between an insulator and a semi-conductor when a gate voltage is applied; see Fig.1. The gas is confined to a domain $\Omega$ in the $xy$-plane. A uniform magnetic field $\vec{B}_0\perp\Omega$ is applied to the system. The so-called filling factor
$$\nu:= \frac{n}{e\vert \vec{B_0} \vert/h\,c }\,,$$
where $n$ is the particle density of the 2D EG, $e$ is the elementary electric charge and $c$ is the speed of light, is chosen such that the longitudinal resistance, $R_L$, of the gas vanishes. It is known, experimentally, that such filling factors exist, and these findings are supported by numerical simulations. But there are essentially no analytical results that enable us to understand why, at certain values of $\nu$, the longitudinal resistance vanishes (or, put differently, why there is a strictly positive mobility gap above the groundstate energy). Here we will not consider this difficult problem, but see, e.g., \cite{Morf}.\footnote{I refer the reader to \cite{Prange} and \cite{Halper}, and references given there, for much useful information on the quantum Hall effect.} Rather, we \textit{assume} that, at certain values of the filling factor, $R_L$ vanishes and then derive non-trivial, mathematically precise consequences of this assumption.
\begin{center}
\includegraphics[width=10cm, height=6cm]{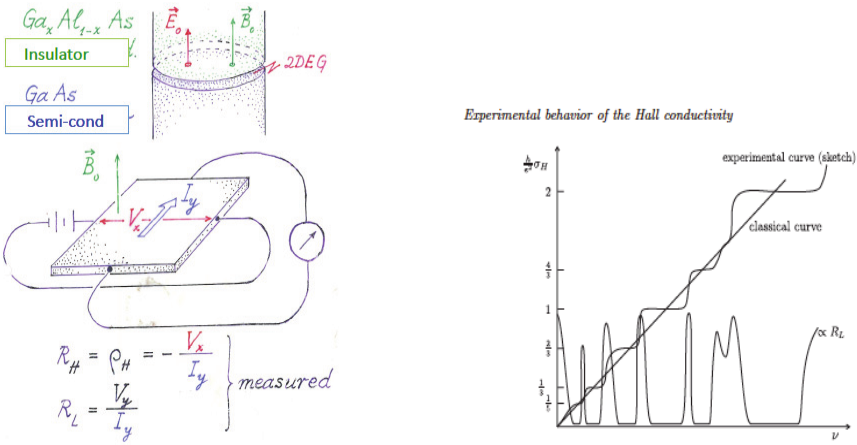}
\end{center}
{\small{\textit{Fig.1 -- Caption:} Schematic representation of the experimental setup and of experimental results on the QHE}\\
\textit{Observations:} $R_{L}=0 \leftrightarrow (\nu,\sigma_{H}) \in$ plateau; plateau heights $\in \frac{e^{2}}{h}\mathbb{Q}$; the cleaner the sample, the more numerous are the observed plateaux and the narrower they are; if $\frac{h}{e^2}\sigma_{H} \notin \mathbb{Z}$ there appear to exist fractional electric charges.\\
\textit{Applications:} Metrology; novel computer memories; topological quantum computing (?) using quasi-particles with non-abelian braid statistics \cite{Froh}}\\

We first study the response of the 2D EG to small perturbations in the external electromagnetic field, $\vec{E}\|\Omega$ and
$\vec{B}\perp\Omega$. We set
$$\vec{B}^{tot}=\vec{B}_0+\vec{B}, \quad B:=|\vec{B}|, \quad \underline{E}:=(E_1,E_2).$$
Note that, as long as electron spin is neglected, only the components, $\underline{E}$, of the electric field parallel to the sample plane and the component, $B$, of the magnetic field perpendicular to the sample plane enter the laws of motion of  electrons in the domain $\Omega$. The field tensor of the electromagnetic field on the three-dimensional space-time of the 2D EG is then given by
\[
{ F:=\left(\begin{array}{c@{\quad }c@{\quad }c}
0 &E_1 &E_2\\
-E_1 &0 &-B\\
-E_2 &B &0
\end{array}\right).}
\]

When interpreted as a 2-form, $F$ is the exterior derivative, d$A$, of a \mbox{1-form}, $A=\sum A_{\mu}\text{d}x^{\mu}$, where $x=(x^{0}\equiv t, \underline{x}),\, \underline{x}=(x^1,x^2) \in \mathbb{R}^{2},$ and $A$ is the electromagnetic vector potential.
\section*{Electrodynamics of 2D incompressible electron gases}

The electric current density in the 2D EG is defined by
\[
j^\mu(x)=\langle J^\mu(x)\rangle_A,\quad\mu=0,1,2,
\]
where $J^{\mu}(x)$ is the quantum-mechanical vector current density -- an operator-valued distribtuion -- and $\langle (\cdot) \rangle_{A}$ is the state of the EG in the presence of an external electromagnetic field with vector potential $A$.\\
We begin by summarizing the basic equations governing the electrodynamics in the bulk of a 2D EG, which, for 
$\underline{E}=0$ and $B=0$, is assumed to be \textit{incompressible}, i.e., $R_{L} =0$.\\

(I) {\bf{Hall's Law}} (assuming that $R_{L}=0$)
\begin{equation}\label{Hall}   %1
\underline{j}(x)=\sigma_H(\underline{E}(x))^*,
\end{equation}
where $\sigma_H$ is the Hall conductivity, and $\underline{E}^{*}$ is the vector arising from counter-clockwise rotation of 
$\underline{E}$ in the ($xy$-) plane of the sample through an angle of $90^{\circ}$.
This equation shows that space reflections in lines, $P$, and time reversal, $T$, are \textit{not} symmetries of the gas, i.e., are broken, which is, of course, a consequence of the presence of the external magnetic field $\vec{B}_{0} \perp \Omega$.\\

(II) {\bf{Charge conservation}}
\begin{equation} \label{conteq}  %2
\frac{\partial}{\partial t}\rho(x)+\underline{\nabla}\cdot\underline{j}(x)=0.
\end{equation}

(III) {\bf{Faraday's induction law}}
\begin{equation}\label{Faraday}   %3
\frac{\partial}{\partial t} B^{tot}_3+\underline{\nabla}\wedge\underline{E}(x)=0.
\end{equation}
It follows that
\begin{equation} \label{4}  %4
\frac{\partial\rho}{\partial t}\mathop{=}\limits^{(2)}-\underline{\nabla}
\cdot\underline{j}\mathop{=}\limits^{(1)}-\sigma_H\underline{\nabla}\wedge\underline{E}
\mathop{=}\limits^{(3)}\sigma_H\frac{\partial B}{\partial t}.
\end{equation}\\
Integrating Eq. \eqref{4} in the time variable $t$, with integration constants chosen as follows
$$
j^0(x):=\rho(x)+e\cdot n,\quad B(x)=\big(\vec{B}^{tot}(x) - \vec{B}_0\big)\cdot \vec{e}_{3}\,,
$$
where $\vec{e}_{3}$ is the unit vector in the $z$-direction orthogonal to $\Omega$, we find the\\
\vspace{0.1cm}

(IV) {\bf{Chern-Simons Gauss law}}
\begin{equation}\label{CS-Gauss}   %5
j^0(x)=\sigma_HB(x)\,.
\end{equation}
\noindent
Combination of laws \eqref{Hall} and \eqref{CS-Gauss} yields
\begin{equation}\label{6}   %6
\framebox{
\hbox{$j^\mu(x)=\frac{1}{2}\sigma_H~\varepsilon^{\mu\nu\lambda}F_{\nu\lambda}(x)$}
}
\end{equation}
Next, when combining laws (II), (III) and Eq. \eqref{6} we seemingly conclude that
\begin{equation}\label{7}   %7
0\mathop{=}\limits^{(2)}\partial_\mu j^\mu
\mathop{=}\limits^{(3),(6)} \frac{1}{2}\varepsilon^{\mu\nu\lambda}(\partial_\mu\sigma_H)F_{\nu\lambda}\ne0,
\end{equation}
wherever $\sigma_H\ne const.$, e.g., at $\partial\Omega$. We seem to arrive at a contradiction. Well, actually, there isn't any contradiction! The point is that the current density $j^\mu$ appearing in Eq. \eqref{6} is the 
{\it bulk} current density, $(j^\mu_{bulk})$, which is \textit{not} the conserved {\it total} electric 
current density, $j^{\mu}_{tot}$:
\begin{equation}\label{8}  %8
j^\mu_{tot}=j^\mu_{bulk}+j^\mu_{edge}, \qquad \text{with  }\,\,\, \partial_{\mu}j^{\mu}_{tot}=0,
\end{equation}
where $j^{\mu}_{edge}$ is the ``edge current density'', a current distribution with support on all lines in $\Omega$ across which the value of 
$\sigma_{H}$ jumps; in particular, $\text{supp}(j^{\mu}_{edge})$ contains the boundary/edge, $\partial \Omega$, of the gas.
In general, only $j^{\mu}_{tot}$ is conserved, but

$$\partial_{\mu} j^{\mu}_{bulk}\,, \partial_{\mu}j^{\mu}_{edge} \not=0,$$
as follows from Eqs. \eqref{7} and \eqref{8}.

\section*{Anomalous chiral edge currents}

The edge current density is defined in such a way that
\[
{\rm supp}~j^\mu_{edge}={\rm supp}(\underline{\nabla}\sigma_H)\supseteq\partial\Omega,\quad 
\underline{j}_{edge}\perp\underline{\nabla}\sigma_H.
\]
One may view it as an instance of {\it ``holography''} that, on supp($\underline{\nabla}\sigma_H$),
\begin{equation} \label{9} %9
\framebox{
\hbox{$\partial_\mu j^\mu_{edge}\mathop{=}\limits^{(8)}
-\partial_\mu j^\mu_{bulk}|_{{\rm supp}(\underline{\nabla}\sigma_H)}
\mathop{=}\limits^{(6)} -\sigma_HE_{\|}|_{{\rm supp}(\underline{\nabla}\sigma_H)}\,,$}
}
\end{equation}
where $E_{\Vert}$ is the component of the electric field $\vec{E}\vert_{{\rm supp}(\underline{\nabla}\sigma_H)}$ perpendicular to $\underline{\nabla} \sigma_{H}$.
Eq. \eqref{9} is the chiral anomaly in $1+1$ dimensions!

The edge current, $j^\mu_{edge}\equiv j^\mu_{\ell/r}$, is an anomalous chiral current in $1+1$ dimensions: At the edge of the sample, the Lorentz force acting on the electrons must be cancelled by the force confining the electrons to the interior of the domain $\Omega$. Hence
$$
\frac{e}{c} \vert B^{tot}v_{\|} \vert= \vert(\underline{\nabla} V_{edge}) \vert, 
$$
where $v_{\Vert}$ is the propagation speed of an electron moving along $\partial \Omega$, and $V_{edge}$ is the potential of the force confining the electrons to the interior of $\Omega$. The chiral nature of $j^{\mu}_{egde}$ becomes obvious from Figure 2.\\
\begin{center}
\includegraphics[width=6.5cm, height=3.5cm]{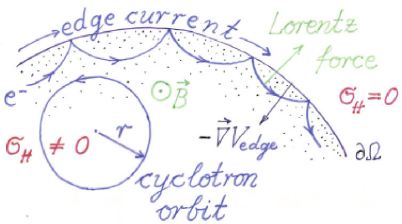}
\end{center}
{\small{\textit{Fig.2 -- Caption:} Skipping orbits of electrons moving along the boundary/edge of a 2D electron gas confined to a disk -- electrons near $\partial \Omega$ perform a chiral motion}}\\

Classically, the orbits of electrons moving close to the edge $\partial \Omega$ of the sample are so-called ``skipping orbits'' which describe a chiral motion of electrons, the chirality being determined by the direction of $\vec{B}_0$.

\noindent
There is an analogous phenomenon in classical physics, namely in the theory of hurricanes:
Under the replacements
$$\vec{B}\to\vec{\omega}_{\rm earth} \text{   (angular velocity) },\quad {\rm Lorentz~force}\to
{\rm Coriolis~force}, $$
$$V_{edge}\to {\rm air~pressure},$$
the theory of charge transport in a 2D EG close to the edge of the sample is mapped to the theory of air-transport in a hurricane near the surface of the Earth. Actually, one may view a hurricane as an example of a ``Hall effect'' in a rotating gas. It is expected that a quantized Hall effect is encountered in a rapidly rotating layer of a superfluid Bose gas; \mbox{(see last subsection of Sect. 1).} \\
Recall that the chiral anomaly in $(1+1)$D says that (for $j^{\mu}_{edge}= j^{\mu}_{\ell}$)
\begin{equation}\label{CA-edge}   %10
\partial_\mu j^\mu_{edge}=-\frac{1}{h}\left(\sum_{{\rm\,\alpha}} q^2_\alpha\right)E_{\|}\quad 
\mathop{\Rightarrow}\limits^{\eqref{9}}\quad
\framebox{
\hbox{$\sigma_H=\dfrac{1}{h}\displaystyle\sum_\alpha q^2_\alpha$\,,}
}
\end{equation}
where $q_{\alpha}=e\widetilde{Q}_\alpha$ is the electric charge of an edge current corresponding to a 
counterclockwise-chiral edge mode $\alpha$; (similar contributions come from clockwise 
modes, but with a \textit{reversed sign} on the right side in Eq. \eqref{CA-edge}). Equation \eqref{CA-edge} yields a precise expression for the Hall conductivity 
$\sigma_{H}$ in terms of charges of chiral currents circulating at the edge of $\Omega$. The integer quantum Hall effect is observed if the charge quantum numbers $\widetilde{Q}_{\alpha}=\pm 1,$ for all $\alpha$. The observation of a system in a \textit{fractional} quantum Hall state then implies that there are currents carried by chiral modes with \textit{fractional} charge circulating at the edge of the system. \\
Apparently, starting from the basic equations of the electrodynamics of a 2D EG in an external electromagnetic field, we are led to  rediscover Halperin's edge currents \cite{Ha}; (see also \cite{FGW}).

\section*{Edge- and bulk effective actions}

Eq. \eqref{CA-edge} shows that if $\sigma_H\notin \frac{e^2}{h}{\Bbb Z}$ then there must exist \textit{fractionally
charged} modes (quasi-particles) propagating along supp$(\underline{\nabla}\sigma_H)$ and, in particular, along $\partial \Omega$.
The quantum-mechanical chiral edge current density $J^\mu_{edge}$ turns out to generate a $U(1)$-current algebra that can be described in terms of free massless chiral Bose fields, $\varphi^{\alpha}$. The Green functions of $J^\mu_{edge}$ can be derived from the anomalous 
chiral action, $\Gamma_{\partial\Omega\times{\Bbb R}}(A_{\|})$ in $1+1$ dimensions -- 
see Eq. \eqref{12}, below -- where $A_{\|}$ is the
restriction of the vector potential A to the boundary, $\partial\Omega\times{\mathbb{ R}}$, of the 
space-time of the 2D EG. The action 
$\Gamma_{\partial\Omega\times{\Bbb R}}(A_{\|})$ is anomalous in the sense that it is \textit{not} invariant under gauge transformations of $A_{\|}$.

The anomaly of $\Gamma_{\partial\Omega\times{\Bbb R}}(A_{\|})$ is a consequence of the fact 
that the chiral edge current $J^\mu_{edge}$ is {\it not} conserved, i.e., is anomalous. The anomaly of the edge current
must be cancelled by the anomaly of the bulk current $J^{\mu}_{bulk}$; see Eq. \eqref{9}. The generating function of the Green functions of the bulk current density is the \textit{bulk effective action}, $S_{\Omega\times{\Bbb R}}(A)$, which we will determine presently. This action is not invariant under gauge transformations of $A$ either. Its variation under gauge transformations of $A$ must be cancelled by the gauge variation of an appropriate multiple of the anomalous chiral action 
$\Gamma_{\partial\Omega\times{\Bbb R}}(A_{\|})$.

Next, we derive the expression for the bulk effective action. By the definition of effective actions, the bulk current is given by the variational derivative of  $S_{\Omega\times{\Bbb R}}(A)$ with respect to $A$. Thus,
\begin{eqnarray*}
j^\mu_{bulk}(x) &=&\langle J^{\mu}_{bulk}(x)\rangle_A
\equiv\frac{\delta S_{\Omega\times{\Bbb R}}(A)}{\delta A_\mu(x)}\\[5pt]
&\mathop{=}\limits^{\eqref{6}}& \frac{1}{2}\sigma_H\varepsilon^{\mu\nu\lambda}F_{\nu\lambda}(x),\quad 
x\notin\partial\Omega\times{\Bbb R}\,.
\end{eqnarray*}
It follows that
\begin{equation}\label{11}   %11
\boxed{
S_{\Omega \times \mathbb{R}}(A)=\frac{\sigma_H}{4} \int_{\Omega \times \mathbb{R}} A \wedge \text{d}A 
}
\end{equation}
Note that the action $S_{\Omega \times \mathbb{R}}(A)$ is \textit{not} invariant under gauge transformations of $A$ \textit{not} vanishing at the boundary, $\partial \Omega \times \mathbb{R}$, of the space-time of the sample. 
The \textit{total effective action} is obtained by adding to the anomalous Chern-Simons action in Eq. \eqref{11} the anomalous chiral action $\Gamma_{\partial\Omega\times{\Bbb R}}(A_{\|})$ (see Eq. \eqref{12}, below), multiplied by a coefficient chosen in such a way that the total action is gauge-invariant. In fact, the coefficient in front of 
$\Gamma_{\partial\Omega\times{\Bbb R}}(A_{\|})$ must be equal to $\sigma_{H}$! This observation implies that the \textit{bulk definition} of the Hall conductivity coincides with its \textit{edge definition}; a conclusion that first appeared in \cite{FK, FS}.

Finally, we present an expression for the anomalous chiral action $\Gamma_{\partial\Omega\times{\Bbb R}}$ of a single chiral matter field on $\partial \Omega \times \mathbb{R}$ coupled to a vector potential $a$:
\begin{equation}\label{12}
\Gamma_{\partial\Omega\times{\Bbb R}}(a):=\frac{1}{2}\int_{\partial\Omega\times{\Bbb R}}
\left[a_+a_--2a_\pm\frac{\partial^2_\mp}{\Box}a_\pm\right]{\rm d}^2u,
\end{equation}
where $a=a_+{\rm d}u^++a_-{\rm d}u^-\equiv A_{\|}$, and $u^{+}, u^{-}$ are ``light-cone coordinates'' on the boundary
$\partial \Omega \times \mathbb{R}$ of the space-time of the sample.

\textit{Exercise:} Check that the anomaly of the (bulk) Chern-Simons action $S_{\Omega \times \mathbb{R}}(A)$ in \eqref{11}, which is a boundary term, is cancelled by the one of $\sigma_{H}\,\Gamma_{\partial\Omega\times{\Bbb R}}(A_{\|})$.

Whatever we have said about 2D electron gases in a homogeneous external magnetic 
field exhibiting the quantum Hall effect can be extended to so-called \textit{Chern insulators} \cite{Hald} (see also \cite{FK}), which have 
the property that reflections in lines (``parity'') and time reversal are broken {\it even} in the absence of 
an external magnetic field, (e.g., because of magnetic impurities in the bulk). An experimental realization of Haldane's model \cite{Hald} has been described in \cite{Esslinger}. The 
low-energy physics of quasi-particles in the bulk of a Chern insulator may resemble 
the one of two-component relativistic Dirac fermions coupled to the electromagnetic 
vector potential, with an effective action given by (11), with $A=A_{\rm tot}$, and 
$\sigma_H=e^2/2h$ ($=$ Chern number of a certain vector bundle of Dirac fermion wave 
functions over the Brillouin zone ${\Bbb T}^2$).

\section*{Classification of ``abelian'' Hall fluids and -Chern insulators}

Next, I sketch a general classification \cite{FT} of 2D insulators with broken time reversal and 
parity in topologically protected states (i.e., groundstates of many-body Hamiltonians with a strictly positive mobility gap above the goundstate energy) exhibiting quasi-particles with \textit{abelian 
braid statistics} \cite{Froh}; (``non-abelian states'' are discussed in \cite{FPSW}). Explicit results will be limited to 
Hall fractions in the interval $(0,1]$.

In the following, $J_{tot}$ denotes the {\it total} electric current density (bulk $+$ edge), 
which is conserved, i.e., $\partial_\mu J^{\mu}_{tot}=0$. We consider a general ansatz: 

\begin{equation*}
J_{tot}=\sum^M_{\alpha=1} Q_\alpha J_\alpha,
\end{equation*}
where the densities $J_\alpha$ are \textit{separately conserved} current densities corresponding to 
different quasi-particle species, and the coefficients $Q_\alpha\in{\Bbb R}$ are 
``charges''. (In the simple case of the integer quantum Hall effect for non-interacting electrons, $\alpha$ labels filled Landau levels of which there are $M$, and $Q_{\alpha}=1, \forall \alpha$.) On a three-dimensional, simply connected space-time 
$\Lambda=\Omega\times{\Bbb R}$, a conserved 
current density $J$ can be derived from a vector potential: If $\mathcal{J}$ denotes the 
2-form dual to $J$ then conservation of $J$ implies that ${\rm d}\mathcal{J}=0$, and hence
\[
\mathcal{J}={\rm d}B,
\]
by Poincar\'{e}'s lemma, where the 1-form $B$ is the vector potential of $J$ and is determined by the current 
density up to the gradient of a scalar function $\beta$; i.e., $B$ and $B+{\rm d}\beta$ yield the 
same current density $J$.

\section*{Chern-Simons effective action of conserved currents}

Henceforth we use units where $\frac{e^2}{h}=1$. For a 2D insulator (i.e., a state of matter protected by a mobility gap above the groundstate energy), the field 
theory of the conserved currents $(J_\alpha)^{M}_{\alpha=1}$ in the limit of very 
large distance scales and very low frequencies must be \textit{topological}: If parity and time reversal are broken 
the ``most relevant'' contribution to the action of the currents $J_{\alpha}$ is the Chern-Simons action
\begin{equation}\label{13}
S_\Lambda(\underline{B}, A):=\sum^{M}_{\alpha=1}\int_\Lambda 
\left\{ B_\alpha\wedge{\rm d}B_\alpha+A\wedge Q_\alpha{\rm d}B_\alpha\right\}
+{\rm boundary~terms},
\end{equation}
where $A$ is the electromagnetic vector potential, and the boundary terms 
must be added to cancel the anomalies of the Chern-Simons terms ($1^{st}$ term on right side) in \eqref{13} under the gauge transformations $B\to B+{\rm d}\beta$ and $A\to A+{\rm d}\chi$.

Carrying out the Gaussian functional integrals over the fields $\big(B_{\alpha}\big)_{\alpha=1}^{M}$, we find that
\begin{equation}\label{14}
\int\exp(iS_\Lambda(\underline{B}, A))\prod^{M}_{\alpha=1}{\cal D}B_\alpha
=\exp\left(i\frac{\sigma_H}{4}\int_\Lambda A\wedge{\rm d}A
+\sigma_H\Gamma_\Lambda(A_{\|})\right),
\end{equation}
where
\begin{equation}\label{15}
\sigma_H=\sum^{M}_{\alpha=1}Q^2_\alpha
\end{equation}

Physical excitations (states) of a topological Chern-Simons theory are labelled by networks of Wilson lines contained in the half-space
 $\Lambda_{-} = \Omega \times \mathbb{R_{-}}$ corresponding to negative values of the time variable and ending in points of the spatial surface \mbox{$\Omega \times \lbrace 0 \rbrace$.} 
 Scalar products of physical states of the Chern-Simons theory with action given by \eqref{13} can be calculated
by inserting into the Gaussian functional integral on the left side of Eq. \eqref{14} networks of Wilson lines contained in 
$\Lambda_{-}$ hooked up to reflected networks of Wilson lines, contained in 
$\Lambda_{+}=\Omega \times \mathbb{R}_{+}$, at common intersection points in $\Omega \times \lbrace 0 \rbrace$, so as to yield a network of \textit{gauge-invariant} Wilson networks/loops. The Hamiltonian of Chern-Simons theory vanishes;  bulk excitations are static, reflecting the feature that there is a positive mobility gap above the groundstate energy.

\section*{Classification of 2D ``abelian'' topological insulators with 
broken time reversal -- bulk degrees of freedom}
The operator measuring the electric charge of an excitation of the Chern-Simons theory with action given by \eqref{13}
contained in a region ${\cal O}$ of space, $\Omega \times \lbrace 0 \rbrace$, is given by 
\[
Q_{\cal O}=\int_{\cal O} J^0(x){\rm d}^2x=\sum^{M}_{\alpha=1} Q_\alpha\int_{\partial {\cal O}} B_\alpha\,\,.
\]
The electric charge deposited inside the domain $\mathcal{O}$ by excitations corresponding to a network of Wilson lines can be inferred by inserting, besides the Wilson network (hooked up to its reflection at $\Omega\times \lbrace0\rbrace$), a Wilson loop 
$$\text{exp}\big(i\,t \sum_{\alpha=1}^{M} Q_\alpha \int_{\partial \mathcal {O}} B_{\alpha} \big), \quad t\in \mathbb{R},$$ 
into the functional integral; (see Eq. \eqref{14}). If a network of Wilson lines creates an excitation describing $n$ 
electrons or holes located inside ${\cal O}$ from the ground state of the insulator then its electric 
charge, as measured by the operator $Q_{\cal O}$, is given by $-n+2k$, $k=1,\ldots,n$, where $k$ is the number of holes. If $n$ is odd this excitation must have Fermi-Dirac statistics, if $n$ is even it must have 
Bose-Einstein statistics. This \textit{relation between the electric charge of an excitation 
and its statistics} implies that the $M$-tuples of quantum numbers assigned to lines in Wilson networks creating 
multi-electron-hole excitations must be sites of an \textit{odd-integral lattice}, $\Gamma$, of 
rank $M$, and that the ``co-vector'' $Q:=(Q_1,\ldots,Q_M)$ must belong to the dual lattice $\Gamma^*$, (in fact, $Q$ must be a so-called ``visible vector'' in $\Gamma^{*}$); see \cite{FT, FST}. Hence 
$$\sigma_H=\sum_{\alpha=1,\ldots,M}Q^2_\alpha \in \mathbb{Q}\,,$$ 
i.e., the Hall conductivity $\sigma_{H}$ is a \textit{rational multiple} of $\frac{e^{2}}{h}$. It takes some more effort to show that the so-called Witt sublattice of $\Gamma$ is an $A$-, $D$- or $E$- root lattice; see \cite{FT, FST}. 

\section*{Classification of 2D ``abelian'' topological insulators with 
broken time reversal -- edge degrees of freedom}

The chiral anomaly in the form of Eq. (10) reflects the feature that there are several (namely $N$) species of gapless quasi-particles 
propagating either clockwise or anti-clockwise along the edge $\partial \Omega$ of the 2D EG. The dynamics of these quasi-particles is described by $N$ massless chiral scalar Bose fields 
$\{\varphi^\alpha\}^N_{\alpha=1}$ with propagation speeds $\{v_\alpha\}^N_{\alpha=1}$; (we choose a basis, 
$\big(\varphi^{\alpha}\big)_{\alpha=1}^{N}$, in field space such that the velocity matrix $\big(v_{\alpha}\big)_{\alpha=1}^{N}$ is \textit{diagonal}). We then find that:

\begin{enumerate}
\item[1.] The chiral electric edge current operator and the Hall conductivity are given by
\[
J^\mu_{edge}=e\sum^N_{\alpha=1} \widetilde{Q}_\alpha\partial^\mu\varphi^\alpha,\quad 
\widetilde{Q}=(\widetilde{Q}_1,\ldots,\widetilde{Q}_N),\quad \text{and   }\,\,\, \sigma_H=\frac{e^2}{h} \widetilde{Q}\cdot \widetilde{Q}^T,
\]
respectively.
\item[2.] Multi-electron/hole states localized along the edge of the 2D EG are created by vertex operators
\begin{equation}\label{16}  %12
:\exp i\left(\sum^N_{\alpha=1} q_{\alpha}\varphi^{\alpha}\right):\,,\quad 
q=\left(\begin{array}{c}
q_1\\
\vdots\\
q_N
\end{array}
\right)\in\widetilde{\Gamma}, 
\end{equation}
where the double colons indicate Wick ordering, and $\widetilde{\Gamma}$ is a lattice. The electric charge carried by the vertex operator in \eqref{16} is given by
$$e\, \widetilde{Q}\cdot q\,.$$ 
By the same reasoning process as above, the relation between the charge of an edge excitation created by the vertex operator in \eqref{16} and its statistics 
implies that the ``quantum numbers'' $q$ must correspond to sites in an \textit{odd-integral lattice}, $\widetilde{\Gamma}$, of 
rank $N$. The Witt sublattice of $\widetilde{\Gamma}$ is an $A$-, $D$- or $E$- root lattice and gives rise to an 
$\hat{A}$-, $\hat{D}$- or $\hat{E}$- Kac-Moody current algebra at level 1 acting as a quantum symmetry on the edge states of the 2D electron gas. The matrix 
$$q^{*}:= \big(q^{*\,k}_{\alpha} \big)_{\alpha, k\, = 1,\dots, N}$$
where, for all $k$, the vector $q^{*\,k}$ determines a vertex operator creating a one-electron state propagating along the edge of the 2D EG, is a close analogue of the Cabibbo-Kobayashi-Maskawa matrix of the weak sector of the Standard Model. 

We conclude that:
\item[3.] The classifying data for the edge degrees of freedom of a 2D EG exhibiting the quantum Hall effect are given by
\begin{equation}\label{classdata}
\{\widetilde{\Gamma};~ \widetilde{Q}\in\widetilde{\Gamma}^* (\hbox{``visible''});\,q^{*}; ~ v=(v_\alpha)^N_{\alpha=1}\}.
\end{equation}
The charged quasi-particles predicted by these data exhibit \textit{abelian braid statistics} and transform under $\hat{A}$-, 
$\hat{D}$- or $\hat{E}$- Kac-Moody symmetries at level 1.
\end{enumerate}

For ``ideal'' Hall insulators, one expects that $\widetilde{\Gamma} = \Gamma$ and $\widetilde{Q}=Q$; (see previous subsection on bulk degrees of freedom). But, in general, the two lattices $\Gamma$  and $\widetilde{\Gamma}$ need not be  identical. This can happen when the edge of a 2D incompressible EG has a layered structure. However, in any case,
the constraint that the edge conductivity $\frac{e^{2}}{h} \widetilde{Q}\cdot \widetilde{Q}^{T}$ is equal to the bulk Hall conductivity 
$\frac{e^{2}}{h} Q\cdot Q^{T}$ must \textit{always} be satisfied, as required by gauge invariance.

A (partial) classification of 2D electron gases with quasi-particles exhibiting \textit{non-abelian} braid statistics and quantized Hall conductivity has been presented in \cite{FPSW}.

It is an interesting, partly open problem to classify egde degrees of freedom localized on interfacial lines at which two different electron gases, with non-vanishing Hall conductivities $\sigma^{(1)}_{H} \not= \sigma^{(2)}_{H} \not=0$, join.

\section*{Success of classification, comparison with data}

In \cite{FT, FST}, we have given a list of ``physically plausible'' odd-integral lattices $\Gamma$ and visible vectors 
$Q\in\Gamma^*$ with the property that $0< (\frac{e^2}{h})^{-1}\sigma_H \leq 1$. We write
$$(\frac{e^{2}}{h})^{-1} \sigma_{H} = \frac{n_H}{d_H}, $$
where $n_H$ and $d_H$ are relatively prime integers, with $n_H \leq d_H$. Plotting $d_H$ vertically (downwards) and 
$(\frac{e^{2}}{h})^{-1} \sigma_H$ horizontally (to the right), we have obtained the table displayed in Fig. 3, below, comparing data on 2D electron gases exhibiting the quantum Hall effect observed in laboratories with our theoretical predictions.
This figure is taken from a plenary lecture I presented at the 1994 International Congress of Mathematicians in Zurich \cite{Frohl}.\footnote{On this occasion, I was kindly introduced to the audience by Ludvig Faddeev.}
\begin{center}
\includegraphics[width=10cm, height=8.5cm]{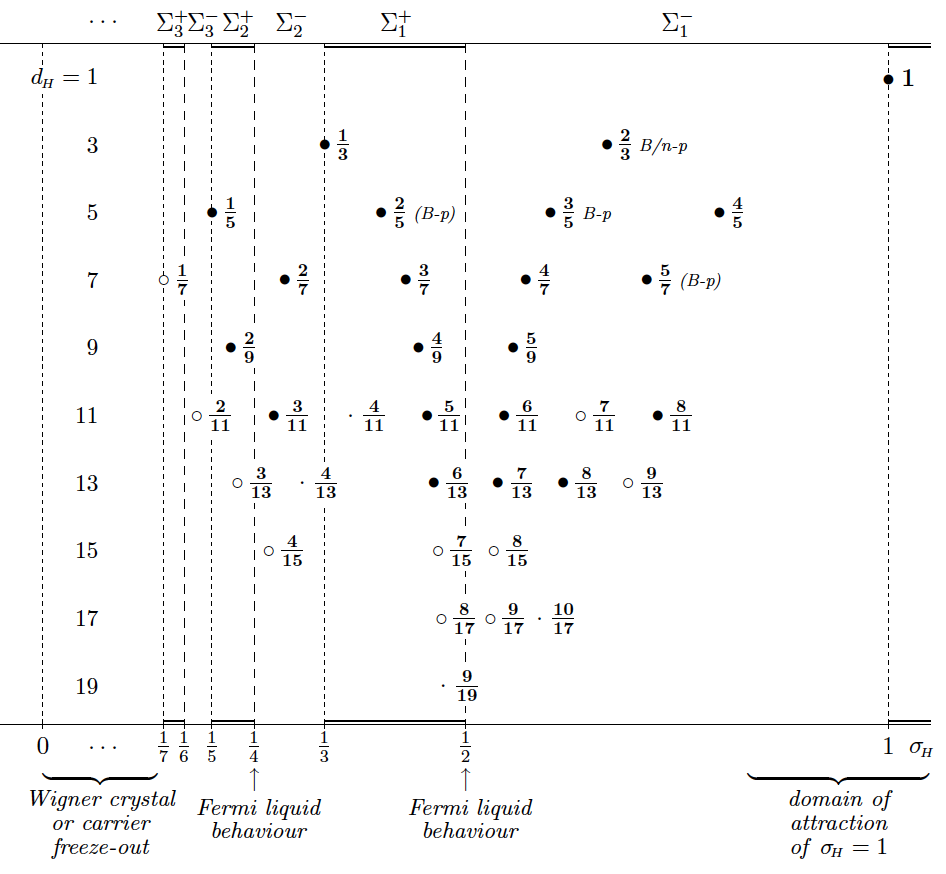}
\end{center}
\small{\textit{Fig.3 -- Caption:} Observed Hall fractions $\sigma_H = n_H/d_H$ in the interval $0<\sigma_H \leq 1$ and their experimental status in single-layer 2D electron gases exhibiting the quantum Hall effect}\\

\textit{Explanation of Figure 3:} Experimentally well established Hall fractions are indicated by a bullet ``$\bullet$''. These are fractions for which the longitudinal resistance, $R_L$, has a minimum very close to $0$ and $R_H$ shows a plateau. Fractions for which $R_L$ has a minimum and $R_H$ shows an inflection are indicated by a circle ``$\circ$''. If there are only weak experimental indications for a Hall insulator state or if the data are controversial the symbol ``$\cdot$'' is used. \textit{All} bullets and circles in this figure, and \mbox{\textit{more} (!),} are predicted theoretically; see \cite{FT, FST}. 

 Hall fractions at which a magnetic field-$(B)$ and/or density-driven transition from one insulator state to another one (at a constant value of $\sigma_H$) has been observed are marked by the letters \mbox{``$B/n-p$'', ``$B-p$''}.
In the interval $[\frac{1}{3},1]= \Sigma^{+}_{1}\cup \Sigma^{-}_{1}$, the agreement between our theory, namely the classification of 2D ``abelian'' Hall insulator states, as presented above, and experimentally observed Hall fractions, including the prediction of transitions at a constant value of $\sigma_H$, is nearly perfect. (Only the fractions $\frac{8}{13}$ and $\frac{10}{17}$ are somewhat uncertain.) Furthermore, there are (abelian and \textit{non-abelian}) states at $\sigma_H=\frac{1}{4}, \frac{1}{2}$ in double-layer electron gases and at $\sigma_H=1$; see \cite{FPSW}. Predictions of plateaux in the intervals $\Sigma^{+}_{p} \cup \Sigma^{-}_{p}, p=2,3,...,$ are related to those in the interval $\Sigma^{+}_{1}\cup \Sigma^{-}_{1}$ by the so-called \textit{``shift map''}; see \cite{FST, FPSW}. For a detailed discussion of the shift map and of the Kac-Moody symmetries of Hall insulators alluded to above, with $\sigma_H<1$ (as indicated in Fig. 3), see \cite{FT, FST, FPSW}.

\section*{Rotating Bose gases, and a duality between photonics and electronics}
We conclude our discussion of Hall- and Chern insulators with some remarks.

\textit{Remark 1:} A discussion of the relevance of the gravitational anomaly and of some new invariants for the theory of the quantum Hall effect (in a spirit somewhat related to the one of this review) can be found in recent papers by P. Wiegmann and collaborators \cite{Wiegmann}.

\textit{Remark 2:} A very rapidly rotating 2D layer of a superfluid Bose gas of atoms (with ``hard-core repulsion'' between atoms) may settle in a state analogous to a Hall insulator. The centrifugal force acting on the bosonic atoms in the gas may be precisely cancelled by a harmonic potential trapping the gas, and the Coriolis force then plays the role of the Lorentz force acting on the electrons in a 2D EG. The energy levels of a bosonic atom precisely correspond to the Landau levels of an electron in a 2D EG. This effect has first been sketched in \cite{FS, FST} and discussed in considerably more detail in \cite{Cooper} -- among other papers. The basic formalism can be found, e.g., in \cite{SP}.

\textit{Remark 3:} In two space dimensions, the electromagnetic vector potential $A$ is \textit{``dual''} to the vector potential 
$B\,\, (:= \sum_{\alpha} Q_{\alpha} B_{\alpha})$ 
of the conserved electromagnetic current density $J$; (see Eqs. (13) and (14)). Under the replacements
$$A\mapsto B, \quad B\mapsto A,$$
\textit{conventional} 2D \textit{insulators} with time-reversal symmetry are mapped to 2D \textit{superconductors}, and conversely, as discussed in \cite{FST}. Furthermore, electronic Hall insulators are mapped to 2D \textit{gapped photonic wave guides} exhibiting extended electromagnetic surface waves. Assuming that time reversal is \textit{not} a symmetry of the system, the leading terms in the action of the electromagnetic field in such a wave guide are given by
 \begin{eqnarray}\label{18}
 S_{\Lambda}^{WG}(A)&=& \frac{1}{2e^{2}} \int_{\Lambda} \lbrace \underline{E}(x) \cdot \varepsilon \underline{E}(x) - \mu^{-1}B^{2}(x) \rbrace \text{d}^{3}x  \nonumber\\
 &+ & \frac{\sigma_H}{4} \int_{\Lambda} A \wedge \text{d}A + \sigma_H\Gamma_\Lambda(A_{\|}),
 \end{eqnarray}
 where $\varepsilon$ is the dielectric tensor of the wave guide and $\mu$ is its magnetic permeability. This is the action functional of \textit{massive} QED in $2+1$ space-time dimensions; see
 \cite{Redlich}. The Chern-Simons term, $\propto \int_{\Lambda} A\wedge \text{d}A$, on the right side of Eq. \eqref{18} arises from coupling the electromagnetic vector potential $A$ to the electric current density $\mathcal{J}=\text{d} B$, as in
 $$ \int_{\Lambda} A\wedge \text{d}B,$$
 assuming that the action functional for the vector potential $B$ of the current density is given, in the simplest instance, by
 $$S_{\Lambda}(B)=\frac{1}{\sigma_{H}} \int_{\Lambda} B\wedge \text{d}B + \text{boundary action} + \text{l.r. terms},$$
as appropriate for an insulator with broken time reversal, (see Eq. \eqref{13}, with $N=1$); the ``boundary action'' (see \eqref{12}) is there to restore gauge invariance, i.e., invariance under $B \mapsto B+\text{d}\beta$, of $S_{\Lambda}(B)$, and ``l.r. terms'' stands for ``less (infrared-) relevant terms'' in the action functional.
 We then find the response equations
 \begin{eqnarray}\label{19}
 \langle F_{\mu\nu}(x) \rangle_{B}&=& \varepsilon_{\mu\nu\rho} \frac{\delta S_{\Lambda}(B)}{\delta B_{\rho}(x)}\nonumber\\
 &=& \sigma_{H}^{-1} \varepsilon_{\mu\nu\rho}j^{\rho}(x)\,,
  \end{eqnarray}
with $\varepsilon_{\mu\nu\rho} j^{\rho}=(\text{d}B)_{\mu\nu}$.\\
 Using arguments similar to those presented in connection with Eqs. \eqref{7} and \eqref{9}, we find that quantized electromagnetic waves can propagate chirally and essentially without dispersion along the edge of such a wave guide. 

\section{Chiral Spin Currents in Planar Time-Reversal Invariant Topological Insulators}
So far, we have not paid attention to electron spin, although there are actually
2D EG exhibiting the fractional quantum Hall effect where electron spin must be taken into account; see \cite{FST}. But it would take us too far off the main line of this paper to discuss
this issue here. \\
In the theory of 2D \textit{time-reversal invariant topological insulators}, which we will sketch next, electron spin undoubtedly  plays a crucial role!
In order to prepare the ground for an account of this theory, we briefly review the Pauli Equation for a non-relativistic spinning electron in an arbitrary external electromagnetic field:

\begin{equation}\label{Pauli}   %13
i\hbar D_0\Psi_t=-\frac{\hbar^2}{2m}\vec{D}\cdot \vec{D}\,\Psi_t,
\end{equation}
where $m$ is the (effective) mass of an electron, 
\[
\Psi_t(\vec{x})=\left(\begin{array}{c}
\psi^\uparrow_t(\vec{x})\\[2pt]
\psi^\downarrow_t(\vec{x})
\end{array}\right)\in L^2({\Bbb R}^3)\otimes{\Bbb C}^2
\]
is a two-component Pauli spinor describing the state of an electron located at the point $\vec{x}$ at time $t$, and the operators $D_0$ and $\vec{D}=(D_1,D_2,D_3)$ are covariant derivatives given by
\begin{equation}\label{D-0}  %14
i\hbar D_0=i\hbar\partial_t\,+\,e\varphi
-\underbrace{\vec{W}_0\cdot\vec{\sigma}}_{{\rm Zeeman~coupling}},\quad 
\vec{W}_0=\mu c^{-1}\vec{B}+\frac{\hbar}{4}\vec{\nabla}\wedge\vec{V},
\end{equation}
where $\varphi$ is the electrostatic potential, $\vec{\sigma}=(\sigma_1,\sigma_2, \sigma_3)$ is the vector of Pauli matrices, $\mu$ is the Bohr magneton, $\vec{B}$ is the magnetic field and $\vec{V}$ is the velocity field generating an incompressible (i.e., $\vec{\nabla}\cdot \vec{V}=0$) classical motion/flow of the system (relative to the inertial laboratory frame).
Furthermore,

\begin{equation}\label{D-k}  %15
\frac{\hbar}{i}D_k=\frac{\hbar}{i}\partial_k+eA_k-m V_k-\vec{W}_k\cdot\vec{\sigma}\,,
\end{equation}
$k=1,2,3,$ where $\vec{A}=(A_1,A_2,A_3)$ is the electromagnetic vector potential, and $\vec{W}_k$ is given by
\begin{equation}\label{spin-orbit}
\vec{W}_k\cdot\vec{\sigma}:=\underbrace{\left[\left(-\tilde{\mu}\vec{E}
+\frac{\hbar}{c^2}\dot{\vec{V}}\right)\wedge\vec{\sigma}\right]_k,}_{{\rm spin}\hbox{-}{\rm orbit~interactions}}\qquad \text{with  }\,\,\tilde\mu=\mu+\frac{e\hbar}{4mc},
\end{equation}
describing spin-orbit interactions and Thomas precession.\footnote{One should actually also include the ``spin connection'' of the sample geometry in the definition of the $SU(2)$-gauge field $\vec{W}_{\mu}$; see \cite{FS}.}
Note that the $2\times 2$ matrices $\vec{W}_0 \cdot \vec{\sigma}$\, and \,$\vec{W}_k \cdot \vec{\sigma},\, k=1,2,3,$\, determine an $SU(2)$-gauge field/connection on the bundle of two-component Pauli spinors over space-time, and that the Pauli equation \eqref{Pauli} obeys $U(1)_{em}\times SU(2)_{spin}$\textit{\,-\,gauge invariance}. \\
For a detailed discussion of the Pauli equation and its gauge invariance, as well as plenty of physical consequences thereof, see \cite{FS, FST}.

\section*{Effective action of a 2D time-reversal invariant topological insulator}
Next, we consider a 2D gas of interacting electrons confined to a domain $\Omega$ in the 
$xy$-plane, assuming that $\vec{B}\perp\Omega$ and $\vec{E}$, $\vec{V}\|\Omega$. Then the 
$SU(2)$- connection, $\vec{W}_\mu$, is given by $W^3_\mu\cdot\sigma_3$, with
$W^K_{\mu}\equiv 0$, $K=1,2$, for $\mu=0,1,2.$

Thus the connection determining parallel transport of the component $\psi^\uparrow$ of a Pauli spinor $\Psi$ is 
given by $a+w$, while parallel transport of $\psi^\downarrow$ is determined by 
$a-w$, with $a_{0}=-e\varphi$, $a_{k}=-eA_{k}+mV_{k}$, and \mbox{$w_\mu=W^3_\mu$}. These connections are \textit{abelian}, i.e., just affect the 
phases of $\psi^{\uparrow}$ and $\psi^{\downarrow}$, respectively. Under time reversal, the components of these connections transform as follows:
\begin{equation} \label{TR}  %16
a_0\to a_0,\quad a_k\to-a_k,\quad but\,\,\,~w_0\to-w_0,\quad w_k\to w_k.
\end{equation}

The term in the effective action of a 2D insulator with the smallest scaling dimension, i.e., the term dominating the large-scale physics, is a Chern-Simons term. If there were only the gauge field $a$, with $w\equiv 0$, or only the 
gauge field $w$, with $a\equiv 0$, a Chern-Simons term would {\it not} be 
invariant under time reversal, and the dominant term in the effective action would be given by
\begin{equation}\label{ins}  %17
S(a)=\int_{\Omega\times \mathbb{R}} {\rm d}^{3}x \,\{\underline{E}\cdot \varepsilon \underline{E}-\mu^{-1}B^2\},
\end{equation}
with $\varepsilon$ the dielectric tensor and $\mu$ the magnetic permeability. This is the effective action of a 
conventional time-reversal invariant insulator.
But, in the presence of electron spin and hence of \textit{two} gauge fields, $a$ and $w$, satisfying \eqref{TR}, the dominant term is a time-reversal invariant combination of two Chern-Simons terms:

\begin{eqnarray}\label{effaction}
S(a,w) &=&\frac{\sigma}{4}\int \{(a+w)\wedge{\rm d}(a+w)-(a-w)\wedge{\rm d}(a-w)\}\nonumber \\[5pt]
&=&\frac{\sigma}{2} \int \{a\wedge{\rm d}w+w\wedge{\rm d}a\}
\end{eqnarray}
This action reproduces an action similar to the one in \eqref{ins} if the connection $w$ is chosen as in \eqref{D-0} and \eqref{D-k}. 
However, the action \eqref{effaction} describes the response of a topological insulator for an \textit{arbitrary} choice  of the gauge field $w_{\mu}=W^{3}_{\mu}$. The gauge fields $a$ and $w$ transform \textit{independently} under gauge transformations, and the Chern-Simons action $S(a,w)$ in Eq. \eqref{effaction} is \textit{anomalous} under gauge transformations of $a\pm w$ \textit{not} vanishing at the boundary, $\partial \Lambda$, of a 2D sample with space-time given by $\Lambda=\Omega\times{\Bbb R}$.
\section*{Edge degrees of freedom: Spin currents}
The anomalous chiral boundary action
\begin{equation}\label{bdaction}
\sigma \lbrace \Gamma((a + w)|_{\|}) - \Gamma((a - w)|_{\|}) \rbrace\,,
\end{equation}
cancels the anomaly of the bulk action $S(a,w)$ in \eqref{effaction}. The two terms in the boundary action are the generating functionals of connected Green functions of two \textit{counter-propagating} chiral electric edge currents, one carried by electrons with ``spin-up'', the other one carried by electrons with ``spin-down'', generating two $U(1)$- current algebras. Thus, whenever the edge degrees of freedom support a current then a net \textit{chiral edge spin-current}, $s^3_{edge}$, circulates at the boundary of the sample.

\section*{Bulk response equations of a 2D time-reversal invariant topological insulator}
 
 From the Chern-Simons action $S(a,w)$ in Eq. \eqref{effaction} one easily derives the following response equations in the bulk of a 2D topological insulator with unbroken time-reversal invariance:
\begin{eqnarray}\label{Bulkresponse} %18
\underline{j}(x)=\sigma(\underline{\nabla}B)^*, \quad \text{and  }\nonumber\\  \vspace{0.1cm} \nonumber\\
s^\mu_3(x)=\dfrac{\delta S(a,w)}{\delta w_\mu(x)}=\sigma\varepsilon^{\mu\nu\lambda}F_{\nu\lambda}(x),
\end{eqnarray}
where $\underline{j}$ is the electric current density in the bulk of the insulator, and $s^{\mu}_{3}$ is the bulk spin-current density, (more precisely its 3-component in spin space); compare to Eqs. \eqref{Hall} and \eqref{6}. By arguments analogous to those in Eqs. \eqref{7} through \eqref{9}, we see that 
Eq. \eqref{Bulkresponse} implies that the spin current $s^{\mu}_{3}$ is anomalous and that its anomaly is cancelled by a chiral egde spin-current.

The Chern-Simons effective action in \eqref{effaction}, the response equations in \eqref{Bulkresponse}, chiral spin currents and the role of spin-orbit interactions in connection with the presence of a non-trivial gauge field $w$ have first been described, among many other effects, in \cite{FS}, in 1993; see also \cite{FST}. 

\section*{Quasi-particles in 2D time-reversal invariant topological insulators}
We should ask what kinds of quasi-particles may induce the effective Chern-Simons actions
\[
S_\pm(a\pm w)=\pm\frac{\sigma}{4}\int (a\pm w)\wedge{\rm d}(a\pm w) \,\,\pm \sigma \Gamma\big((a \pm w)|_{\|}\big)\,,
\]
see Eq. \eqref{effaction}, where ``$+$'' stands for ``spin-up'' and ``$-$'' stands for ``spin-down''. 
It is well known that a two-component relativistic 
Dirac fermion with mass $M>0$ ($M<0$) breaks time-reversal invariance and -- when coupled to an abelian gauge field, $A$ --  induces a Chern-Simons term
\[
\mathop{+}\limits_{(-)}\frac{1}{4\pi}\int A\wedge{\rm d}A
\]
in the effective action; see \cite{Redlich}.
Thus, a 2D time-reversal invariant topological insulator with chiral 
edge spin-currents might exhibit two species of charged quasi-particles in the bulk, 
with one species (spin-up) related to the other one (spin-down) by time reversal, 
each species having two degenerate states per wave vector and mimicking, at small wave vectors, a 2-component 
Dirac fermion. The value of the constant $\sigma$ is then determined!

\section*{Experiments, other approaches, chiral spin liquids}
Materials with properties similar to those predicted by the theory described above have been studied experimentally in the group of L. Molenkamp in W\"urzburg \cite{Molenkamp}. More detailed theoretical aspects have been discussed in \cite{Zhang, GP}. 
However, because of external electromagnetic fields violating the assumption that $\vec{B} \perp \Omega$ and $\vec{E} \Vert \Omega$, and because of magnetic impurities in the material, the property that $\vec{W}_{\mu}=(0,0,W^{3}_{\mu})$ will usually not hold exactly, and hence the spin current $s^{\mu}_{3}$ is \textit{not} conserved; (although $\vec{s}\,^{\mu}$ is always ``\textit{covariantly} conserved'', \cite{FS,FST}). There then exist spin-umklapp processes that cause transitions between the two chiralities of edge currents and hide the quantization of the value of $\sigma$. For these reasons, one cannot expect that real materials exactly reproduce the theoretical predictions of the theory sketched above.

One may, however, expect that there \textit{are} materials with edge currents, $j_{\ell}$ and $j_{r}$, of opposite chirality related to each other by time reversal that do \textit{not} mix. The current $j_{\ell}$ may couple to a chiral abelian gauge field $a_{\ell}$ and the current $j_{r}$ to a gauge field $a_r$. The effective actions of $a_{\ell}$ and $a_{r}$ are given by $(\pm 1)\, \times$ the anomalous action in \eqref{12}. The two gauge fields may have the form 
$$a_{\ell/r} = a \pm w,$$
where $w$ is a Berry connection. The transformation of $a$ and $w$ under time reversal is the one given in \eqref{TR}.
The theory sketched above provides a description of such materials. This theory has the advantageous feature that it is \textit{not} based on a single-particle picture. More recently, $\mathbb{Z}_{2}$\,- topological invariants based on a single-particle picture and characterizing non-interacting time-reversal invariant topological insulators have been studied by Kane and Mele and numerous further authors; see, e.g., \cite{GP} and references given there.

In \cite{FS,FST}, various further 2D systems with somewhat exotic magnetic properties have been described. For example, it has been speculated (see also \cite{FW}) that there may be \textit{chiral spin liquids} with broken time-reversal symmetry
not exhibiting any long-range spin order. The response laws for the spin current, $\vec{s}\,^{\mu}$, of chiral spin liquids can be derived from a non-abelian Chern-Simons action,
$$\frac{k}{4\pi}\int_{\Lambda} \text{tr}\big( w\wedge \text{d}w + \frac{2}{3}w\wedge w \wedge w \big) + \text{bd. terms},$$
where $k$ is an integer, and $w=\sum w_{\mu}\text{d}x^{\mu}$ is an $SU(2)$-gauge field coupling to electron spin, with $w_{\mu}=\vec{W}_{\mu}\cdot \vec{\sigma}$. 

\section{3D Topological Insulators, Axions}

Encouraged by the findings of the last section, we propose to consider insulators 
in three dimensions confined to a region
\[
\Lambda :=\Omega\times{\Bbb R},\quad\Omega \subset \mathbb{R}^{3}, \,\, \partial\Lambda\ne\emptyset
\]
of space-time. We are interested in the general form of the effective action 
describing the response of such materials to turning on an external electromagnetic field. Until 
the mid nineties, it was expected that the effective action of a 3D insulator is always given by
\begin{equation}\label{3DI}  %19
S_\Lambda(A)=\frac12\int_\Lambda {\rm d}t\,{\rm d}^3x\{\vec{E}\cdot\varepsilon\vec{E}
-\vec{B}\cdot\mu^{-1}\vec{B}\},
\end{equation}
where $A$ is the electromagnetic vector potential of the electromagnetic field $(\vec{E}, \vec{B})$, $\varepsilon$ is the dielectric tensor of the material and $\mu$ is 
the magnetic permeability tensor. In $(3+1)$ dimensions, the action in \eqref{3DI} is dimensionless.

 In the seventies, particle 
theorists have taught us that one could add another dimensionless term to the action in \eqref{3DI}, the  
$\theta$-term first introduced in QCD.

\section*{An effective action with a topological term}
We replace the action $S_{\Lambda}(A)$ by
\begin{equation}\label{3DTI}   %20
S_\Lambda(A)\,\rightarrow\, S^{(\theta)}_\Lambda(A):=S_\Lambda(A)+\theta\, I_\Lambda(A), \quad \theta \in [0, 2\pi)\,,
\end{equation}
where $\theta$ is the analogue of the ``vacuum angle'' in particle physcis, and $I_\Lambda$ is a topological term (``instanton number'') given by
\begin{equation} \label{Taction}  %21
\framebox{
\hbox{$(4\pi^2)I_\Lambda(A)=2 \displaystyle\int_\Lambda{\rm d}t\,{\rm d}^3x\,\vec{E}
\cdot\vec{B}=\int_\Lambda F_A\wedge F_A\mathop{=}\limits_{\rm Stokes}\int_{\partial\Lambda} A
\wedge{\rm d} A$}
}
\end{equation}
see \cite{FST, Wang, FW}.
The partition function of a topological insulator in three dimensions whose effective action is given by \eqref{3DTI} and \eqref{Taction} is
\[
\mathcal{Z}^{(\theta)}_{\Lambda}(A)=\exp\big(iS^{(\theta)}_\Lambda(A)\big).
\]

\noindent
In the thermodynamic limit, $\Omega\nearrow{\Bbb R}^3$, $\mathcal{Z}^{(\theta)}_{\Lambda}(A)$
is {\it periodic} in $\theta$ with period $2\pi$ and {\it invariant under time reversal} 
iff
$$\theta=0,\pi\,.$$

\section*{Surface degrees of freedom}

A conventional insulator corresponds to $\theta=0$. For $\theta=\pi$, the partition function
${\cal Z}^{(\theta)}_\Lambda(A)$ has a boundary term given by
\begin{equation}\label{CS-term} %22
\exp\left(\frac{i}{4\pi}\int_{\partial\Lambda} A\wedge{\rm d}A\right),
\end{equation}
see \eqref{Taction}. As mentioned in the last section, this is the leading term in 
the partition function of $(2+1)$-dimensional charged,
``relativistic'', two-component Dirac fermions on $\partial\Lambda$ coupled to an external electromagnetic field $A\vert_{\partial \Lambda}$.  Two species of charged spin-polarized quasi-particles with a conical 
dispersion law mimicking a two-component relativistic Dirac fermion and propagating along the surface of the system
may be encountered in certain 3D topological insulators with two degenerate bands ``communicating'' with each 
other and an effective action given by \eqref{3DTI}, \eqref{Taction}, with $\theta=\pi$. Such systems support surface Hall currents, as is evident from the form of the surface action in \eqref{Taction} and \eqref{CS-term}; cf. Eq. \eqref{11}.
They have been studied experimentally in \cite{Molenkamp2}; see also \cite{Wang}.

\section*{Axionic topological insulators}

One may wonder whether, following ideas of Peccei and Quinn \cite{PQ}, it might make sense to view 
the angle $\theta$ as the (groundstate) 
expectation value of a dynamical field, $\varphi$, and replace the term 
$\theta\,I_\Lambda(A)$ by
\begin{equation}\label{axion-TI}   %23
I_\Lambda(\varphi,A):=\frac{1}{4\pi^2}\int_\Lambda \varphi\,F_A\wedge F_A+S_0(\varphi),
\end{equation}
where $S_0(\varphi)$ is invariant under constant shifts $\varphi\mapsto\varphi+n\pi$, 
$n\in{\Bbb Z}$, and $\varphi$ is a pseudo-scalar field called ``axion field'', (see \cite{FP}).
The effective action given in \eqref{axion-TI} then gives rise to Halperin's 3D Hall effect:
\begin{equation} \label{3D-HE} %24
\framebox{
\hbox{$j=-\dfrac{1}{2\pi^2}{\rm d}\varphi\wedge F_A;\quad {\rm in~particular}\quad 
\vec{j}=\dfrac{1}{2\pi^2}\vec{\nabla}\varphi\times\vec{E}$}
}
\end{equation}
In a crystalline insulator, and for an axion linear in $\vec{x}$, 
$\vec{\nabla}\varphi={\rm cst.}\vec{K}$, where, by invariance under lattice translations, $\vec{K}$ must belong to the dual 
lattice. Hence $\vec{\nabla}\varphi$ is ``quantized''.\footnote{I thank Gregory Moore for having explained this point to me; see \cite{Moore}} It is argued that dynamical axions may emerge 
in certain topological insulators with anti-ferromagnetic 
short-range order and two bands of electron states ``communicating'' with 
each other. The time derivative of $\varphi$ then has the interpretation of a 
chemical potential conjugate to the axial charge: The chiral anomaly tells us that
\begin{equation}\label{4D-CA}  %25
\int {\rm d}\varphi\wedge j_5= - \text{cst.} \int \varphi F_A\wedge F_A + \,\text{terms }\propto \text{ masses }
\end{equation}
The action of Eq. \eqref{axion-TI} has been proposed in \cite{FP} as the effective action describing a higher-dimensional cousin of the Hall effect and has been applied to the problem of unravelling possible origins of intergalactic primordial magnetic fields in the Universe; see also \cite{ACF}. Later it appeared in \cite{Hehl, SCZhang}.

\section*{Mott transition and instabilities in axionic topological insulators}

It is interesting to analyze physical effects observable in putative states of matter described by the effective action
$$S^{tot}_{\Lambda}(A):= S_\Lambda(A)+ I_{\Lambda}(\varphi, A),$$ 
where $S_\Lambda(A)$ is as in \eqref{3DI} and $I_{\Lambda}(\varphi, A)$ as in \eqref{axion-TI}.
Since the action $S_{0}(\varphi)$ on the right side of Eq. \eqref{axion-TI} has been assumed to be periodic under constant shifts 
$\varphi \mapsto \varphi+ n \pi, n \in \mathbb{Z}$, one may expect that, at positive (but low) temperatures,
the bulk of such a state of matter will exhibit \textit{axion domain walls} across which the mean value of $\varphi$ changes by an integer multiple of $\pi$. Recalling the insight described after \eqref{CS-term}, one predicts that gapless, charged two-component Dirac fermions propagate along domain walls across which the mean value of $\varphi$ changes by an odd multiple of $\pi$. We are thus led to predict that the formation of macroscopic axion domain walls gives rise to a ``Mott transition'' from an insulator to a state with non-vanishing bulk conductivity, due to the extended charged surface modes propagating along such domain walls; (see \cite{FW}).

Axion-electrodynamics exhibits some tantalizing \textit{instabilities}. One such instability, triggered by the term 
$I_{\Lambda}(\varphi, A)$, manifests itself in the growth of long-wave-length magnetic fields in the bulk of the system; see \cite{ACF, FP}. A related instability has been described in \cite{OO}: Above a certain critical field strength, $E_{c}$, a constant external electric field, $E\vec{n} \,\,(\text{with }\vert \vec{n}\vert =1 \text{ and } E>E_{c}),$ applied to an axionic topological insulator is \textit{screened}, with the excess field,
$(E-E_{c})\vec{n},$ converted into a \textit{magnetic field}.

Expository articles covering theoretical and experimental aspects of topological insulators can be found, e.g., in \cite{Joel, Franz}.  

\vspace*{2pt}
\section*{Conclusions and outlook}
\begin{itemize}
\item {The Physics of 2D (and 3D) insulators is surprisingly rich and has potential to lead to
promising technological applications. Interesting mathematical techniques -- 
ranging from abstract algebra over the topology of fibre bundles all the way 
to hard analysis -- have non-trivial applications to problems concerning these fascinating states of matter.}

\item {2D electron gases, Bose gases and magnetic materials represent stimulating play 
grounds for experimentalists and theorists alike -- not least because very general 
principles of quantum theory, such as \textit{braid statistics, fractional spin \& 
fractional electric charges, anomalies and their cancellation, current algebra} and \textit{ 
holography} appear to manifest 
themselves in the arena of 2D condensed matter physics. Unfamiliar species of quasi-particles, in particular \textit{two-component Dirac-like fermions} and \textit{Majorana fermions} are encountered in the physics of various exotic two- and three-dimensional systems whose exploration has begun only recently.}

\item{It is interesting to consider higher-dimensional cousins of the QHE and 
of time-reversal invariant topological insulators; see \cite{FP, zilber}. Some of them might be 
relevant in cosmology, e.g., in connection with the generation of intergalactic primordial 
magnetic fields in the Universe. Ideas on cosmology somewhat related to those reviewed in this paper are presently actively pursued.}

\end{itemize}

Many ideas originally developed in particle physics turn out to have fruitful applications in condensed matter physics. Likewise, ideas originating in condensed matter physics and statistical mechanics promise to have very interesting 
applications in cosmology. But that's another story that I cannot go into here. -- In any event, these fields of physics appear to feature highly stimulating exchanges and fruitful interactions, or call for such. Moreover, different mathematical techniques and styles of research have turned out to be useful and to complement each other in productive ways in particle physics, condensed matter physics and cosmology. New insights gained in pure mathematics sometimes have spectacular applications in physics, and conversely. Yet, disappointingly, interactions across certain disciplinary boundaries apparently continue to be weaker than they ought to be, and people in different communities are not seldom reluctant to profit from each other's experiences. A determination to change this belongs to the legacy Ludvig Faddeev has left us.\\

\end{document}